\documentclass[epj]{svjour}
\usepackage{graphics}
\usepackage{eurosym}
\usepackage[usenames]{color}
\usepackage{graphicx}
\usepackage{amsmath}
\usepackage{amsfonts}
\usepackage{amssymb}
\usepackage{mathrsfs}
\usepackage{bm}
\usepackage{verbatim}
\usepackage{ulem}

\newcommand{\beq}{\begin{equation}}
\newcommand{\eeq}{\end{equation}}
\newcommand{\bea}{\begin{eqnarray}}
\newcommand{\eea}{\end{eqnarray}}

\graphicspath{{../}}

\begin{document}
\title{Do we know how to count powers in pionless and pionful effective
field theory?}

\author{C.-J.~Yang
}                     
%
%
\institute{Department of Physics, Chalmers University of Technology, SE-412 96
G\"oteborg, Sweden }
\date{Received: date / Revised version: date}
%
\abstract{
In this article I summarize recent progress in the effective field theory
approach to low energy nuclear systems, with a focus on the power counting
issue. In the pionless sector, where the power counting is quite well
understood at the nucleon-nucleon (NN) level, I discuss some recent
developments toward few- and many-body calculations. In the pionful sector,
I focus on the actively debated issue of power counting in the NN sector and some
recent developments toward a model-independent NN interaction. Finally, the
scenario that the power counting might depend on the number of particles is discussed.
%
} 
\maketitle
\section{Introduction}

The pursuit towards a truly model-independent description of low-energy ($<1$%
GeV) nuclear systems have been carried out through Effective field theory
(EFT) for several decades. In this approach, one first builds the
inter-nucleon interaction through a Lagrangian which captures important
symmetries of QCD at low energy, and then carries out \textit{ab-initio} calculations
based on the resulting interaction to predict nuclear properties.

The main idea of EFT is to build a theory which works within the
momentum scale of interest without knowing or assuming physics in other
places. Therefore, a prerequisite is that physics at the scale of interest
can be separated from unimportant details\footnote{%
Here we define unimportant as both physics that has been integrated-out by
applying regulators and physics (Feynman diagrams) dropped beyond the
applicability of that order.}---which is normally the ultraviolet physics.
If this is the case, then one has at least two momentum scales in the
theory, i.e., the high-energy scale $M_{hi}$ which characterizes our
ignorance of ultraviolet physics, and the low-energy scale $M_{lo}$ which
characterizes the physics of interest. Ideally, after renormalization it is
desirable to arrange physical observables order by order in an expansion of $%
\frac{M_{lo}}{M_{hi}}$. In the case where one adopts a cutoff $\Lambda$ in
the regulator, an observable $\mathcal{O}$ evaluated up to order $n$ can be
expressed as \cite{gries}: 
\begin{align}
&\mathcal{O}_n(M_{lo};\Lambda;M_{hi})=\sum_i^n\left(\frac{M_{lo}}{M_{hi}}\right)^i\mathcal{F}_i(M_{lo};M_{hi})\notag \\ 
&+\operatorname{\mathscr C}_n(\Lambda;M_{lo},M_{hi})\left(\frac{M_{lo}}{M_{hi}}\right)^{n+1},
\label{pc}
\end{align}
where $\mathcal{F}_i$ is a function which includes physics at order $\textit{i}$, so that the first term in the right-hand side of Eq. (\ref{pc}) 
can be improved order by order by calculating loops.
Note that there is no $\Lambda$-dependence in $\mathcal{F}_i$, as it is of higher-order after renormalization.
The residue $\operatorname{\mathscr C}$ is a function of $\Lambda$, $M_{lo}$ and $M_{hi}$ and represents higher order effects which has
not been evaluated. Although the exact form is unknown, the size of $\operatorname{\mathscr C}$ should be of natural size for $%
M_{lo}<M_{hi}$ and $\Lambda>M_{hi}$. The latter condition also ensures that
there is no further cut in the relevant part of physics ($k=0\sim M_{hi}$), so that if the
renormalization is performed correctly, 
$\operatorname{\mathscr C}$ depends on negative power of $\Lambda$. To arrange relevant
terms (from the Lagrangian) and to generate $\mathcal{O}_n$ at each other as close as
possible as described in Eq. (\ref{pc}) requires power counting.

In nuclear systems, the low-energy scale $M_{lo}$ usually contains the center
of mass (c.m.) momentum of the nucleon $p_{cm}$ or some other typical momentum
scale $p_{typ}$ such as the pion mass $m_{\pi}$, i.e., ${p_{cm},p_{typ}}%
\subset M_{lo}$. The breakdown scale $M_{hi}$ depends on the degrees of
freedom included in the theory. For pionless EFT, where the theory includes
only protons and neutrons as degrees of
freedom, the breakdown scale $%
M_{hi}\sim140$ MeV since effects of pion-exchange are not included. For
pionful EFT, where the theory includes nucleons and pions as degrees of
freedom, the estimated breakdown scale ranges from $M_{hi}=600-1000$ MeV
depending on whether one counts the first excitation beyond pions---the $%
\sigma$ or $f_0(600)$---as the breakdown scale or just adopts the value $%
4\pi f_{\pi}\sim 1000$ MeV from chiral perturbation theory\footnote{$f_{\pi}\sim 93$ MeV is the pion decay constant and $4\pi f_{\pi}$
is the suppression comes from extra pion loop in an irreducible diagram.}. When properly
organized, EFT should be able to provide reliable predictions for processes
where the momentum $p_{cm}$ involved is within the breakdown scale.

To properly organize an EFT, in most of the cases, renormalization is
necessary as physics which has been integrated out is absorbed and encoded
in the low energy constants (LECs) associated with contact terms. It is then
of importance to check whether the results after renormalization satisfy
the renormalization group (RG) requirement\footnote{In this work the word ``RG-invariant" refers to
cases where the result converges with respect to $\Lambda$, i.e., the observable can only depend on negative power of $\Lambda$ after renormalization.}. Note that one should not
mix the scale $M_{hi}$ with the cutoff $\Lambda$. The expansion in Eq. (\ref%
{pc}) is valid as long as $M_{lo}<<M_{hi}$ regardless of value of $\Lambda$.

In pionless EFT, where the pionful degrees of
freedom has been integrated out, RG
and the power counting can be checked analytically in the nucleon-nucleon (NN) sector%
 \cite{bira9808007,bira99,pionless,pionlessa,pionlessb,pionlessc,pionlessd,pionlesse,pionless1}. For three-particle systems, the power counting is also
well-studied \cite{pionless3,pionless3aa,pionless3b,pionless3c,pionless3d,pionless3e,pionless3a,pionless3ab,pionless3ac,pionless3cou,pionless3cou2,pionless3cou3,pionless3cou1,unitarity,pionless16,pionless16b,pionlessmatter}. One surprising feature is that a
three-body force is required already at leading order (LO) to prevent the triton
from Thomas collapsing \cite{thomas,thomas2,thomas3,thomas4}. Moreover, a recent study \cite{pionless4}
suggested that, at least for the bosonic systems, a four-body force is required at next-to leading order (NLO).

The investigation of RG and power counting is more involved in the pionful
sector. An analytical solution for the NN-amplitude is already impossible and all
studies must be carried out numerically. Due to this difficulty, a common
approach is to apply power counting at the potential level based on
Weinberg's prescription (WPC) \cite{We90,We91} and then iterate the potential which is
truncated at a certain order in the Schrodinger or Lippmann-Schwinger equations, in order to
obtain the observables. This non-perturbative treatment, though practical
in ab-initio calculations, does not satisfy the RG requirement \cite%
{nogga,Ya09A,Ya09B,ZE12}. As a result, whether RG needs to be
satisfied in the pionful case is still in an ongoing debate. Refs. \cite{ge,ge1,ge1b,ge1c} argued that
adopting a cutoff higher than a certain value (which normally ranges from $%
450\sim 600$ MeV) will cause the ``peratization" of an EFT and generate
meaningless results\footnote{See Ref. \cite{ge} for the meaning of ``peratization" in this context and Refs. \cite{vald19,vald19b,ge_res} for a recent debate of the above issue.}. 
On the other hand, efforts toward building the interaction which satisfies the same criteria as in the pionless case or any other quantum field
theories have been carried out and
resulted in three versions of alternative power counting \cite{Birse,Birseb,Birsec,Valdper,Valdperb,BY,BYb,BYc}%
. All of them treat subleading corrections perturbatively and are able to
generate RG-invariant and reasonable NN amplitudes with respect to those obtained from WPC. 

The rest of this article is organized as follows. Section II provides a
simple overview regarding power counting. Section III reviews power counting
in pionless EFT. Section IV deals with power counting in pionful EFT.
Finally, a summary of current situation regarding power counting in EFT is
given in section V.

\section{What is power counting? How to validate it?}
\label{secpc}
One main ingredient of EFT is the power counting, which tells us how to generate
the final observable order-by-order from a given Lagrangian. Since the EFT expansion
is to be arranged on the final observable as listed in Eq. (\ref{pc}), power counting should
be applied directly to the observable instead of some intermediate
quantities such as the potential---though estimations of those quantities based on naive dimensional analysis (NDA) \cite{howard} often provides a first useful guide in truncating the infinite series. 
This means power counting
is system-dependent. Factors such as the energy scale, number of particles, whether there are bound states or not in the systems
need to enter the power counting. Applying one power counting which works fine in one
system to another could produce completely wrong result if a factor which is originally unimportant becomes important in the new system.

One way to check power counting is to perform a trial and error
procedure as follows. First, one assumes a power counting based on naive
dimensional analysis (NDA) \cite{howard} or other insights and uses it to calculate the
observables order by order. Then, one checks whether the observable at each
order actually matches the assumed power counting. For observables which can
be expressed as a function of momentum, a simple check can be performed
by utilizing the residue-cutoff-dependence as described in Ref.  \cite%
{gries}. In this approach, one generates observables at two different
cutoffs ($\Lambda_1$, $\Lambda_2$) and subtracts the two results with each other. From Eq. %
(\ref{pc}) one reaches: 
\begin{align}
&\frac{\mathcal{O}_n(p_{cm},p_{typ};\Lambda_1)-\mathcal{O}_n(p_{cm},p_{typ};%
\Lambda_2)}{\mathcal{O}_n(p_{cm},p_{typ};\Lambda_1)}= \notag \\
&\left(\frac{%
p_{cm},p_{typ}}{M_{hi}}\right)^{n+1}\frac{\operatorname{\mathscr C}%
_n(\Lambda_1;p_{cm},p_{typ},M_{hi})-\operatorname{\mathscr C}_n(%
\Lambda_2;p_{cm},p_{typ},M_{hi})}{\mathcal{O}_n(%
\Lambda_1;p_{cm},p_{typ},M_{hi})}.  \label{pc2}
\end{align}
The slope of a double-logarithmic plot against ln$(p_{cm})$, i.e., 
\begin{align}
& \ln\left(\frac{\mathcal{O}_n(p_{cm},p_{typ};\Lambda_1)-\mathcal{O}_n(p_{cm},p_{typ};\Lambda_2)}{\mathcal{O}_n(p_{cm},p_{typ};\Lambda_1)}\right) \notag \\
& \sim(n+1)\ln\left(\frac{p_{cm},p_{typ}}{M_{hi}}\right)
 \label{eq2}
\end{align} 
corresponds to the power $n+1$. The above procedure has been carried out in
the neutron-deuteron scattering process to determine the power counting in
the pionless case \cite{gries}, where it is demonstrated that useful
information regarding power counting of the three-body force can be extracted.

Before the above method is proposed, G. P. Lepage \cite{lepage} proposed a similar check by
directly examing the difference between results up to certain order and
the data as a function of the c.m. energy $E_{cm}$. However, information
extracted in this way---the so-called Lepage plot---is not as clean as Eq. %
(\ref{eq2}). Different observables adopted in the renormalization could generate sizable difference not just in the residue $\operatorname{\mathscr C}_n$, but in $\sum_i^n\left(%
\frac{M_{lo}}{M_{hi}}\right)^i\mathcal{F}_i(M_{lo};M_{hi})%
$ also.
Therefore, a direct subtraction of theoretical
result from experimental data could create sizable uncertainty on the final
extracted slope, unless all LECs at different order are fixed in a very
particular way to minimize this uncertainty\footnote{%
Ref. \cite{lepage} fixes all LECs at very low energies.}.

The above two methods require the extraction of power counting to be performed at cutoffs
large enough so that $\left(\frac{M_{lo}}{M_{hi}}\right)>\left(\frac{M_{lo}}{%
\Lambda}\right)$, otherwise effects from the cutoff in the un-converged results
could enter and contaminate the extracted value. Thus, a prerequisite
is that RG needed to be satisfied in the first place.
Meanwhile, one cannot rule out the possibility that under a limited window
of cutoff, a non-RG-invariant theory could generate the
same results as those generated from the correct EFT. To check power counting under $\Lambda<M_{hi}$, some
methods are proposed \cite{Epelmore2,bay}. The simplest check \cite{Epelmore2}
is to examine whether the correction at each order divided by the LO ,
e.g., $\frac{\mathcal{O}_i(p)}{\mathcal{O}_{LO}(p)}$, scales as $\left(\frac{%
p,m_{\pi}}{M_{hi}}\right)^i$. However, one needs to assume a numerical value
of the breakdown scale $M_{hi}$. This simple check can also be plagued by
effects from fitting strategies and terms proportional to $\left(\frac{M_{lo}}{%
\Lambda}\right)^i$. 
A more advanced method given in Refs. \cite{bay,bayb,bayc,bayd,baye,bayf,bayg} utilizes the order-by-order convergence of observable predictions from a given EFT, fit to data, as a diagnostic tool for EFT power-counting.
If the order-by-order observables show clear issues in the statistical diagnostics, it is a sign that the EFT is not converging as expected for a given observable.

\section{Power counting in pionless EFT}

\subsection{NN level}

The Lagrangian of pionless EFT reads \cite{bira9808007}
\begin{align}
&L_{NN}=N^{\dag }(i\partial _{0}+\frac{\overrightarrow{\nabla }^{2}}{2M_N}+...)N \notag \\
&-\frac{1}{2}C_{0}(N^{\dag }NN^{\dag }N)-\frac{1}{8}(C_{2}+C_{2}^{\prime
})\notag \\ 
&[N^{\dag }(\overrightarrow{\nabla } -\overleftarrow{\nabla })N\cdot N^{\dag
}(\overrightarrow{\nabla }-\overleftarrow{\nabla })N-N^{\dag }NN^{\dag }(%
\overrightarrow{\nabla }-\overleftarrow{\nabla })^{2}N] \notag \\ 
&+\frac{1}{4}(C_{2}-C_{2}^{\prime })N^{\dag }N\overrightarrow{\nabla }%
^{2}(N^{\dag }N)+...,\label{la}
\end{align}%
where $M_N$ is the nucleon mass, $N$ is the nucleon field and $C_{2n}^{(\prime
)}$'s are LECs. The above Lagrangian results an interaction in form of%
\begin{equation}
v(p,p^{\prime })=C_{0}+C_{2}(p^{2}+p^{\prime 2})+2C_{2}^{\prime }%
\overrightarrow{p}\cdot \overrightarrow{p}^{\prime }+..., \label{eq3}
\end{equation}
where $p^{(\prime)}$ is the c.m. momentum of the nucleon. 
Note that, unlike the case in QED or QCD, the effective Lagrangian in Eq. (\ref{la}) contains infinitely many terms and cannot be solved exactly. In fact, the fully reliable function of it is to
provide vertices and propagators under certain symmetries. The derivative expansion serves as a motivation for the power counting, but in a less reliable sense,
as the LECs are to be utilized to describe observables and their relative importance 
is system-dependent. Beforehand the power counting is unknown.
To illustrate the idea, two scenarios of the power counting are given below as examples.

The first scenario is the simplest case, 
where every LECs with $2n$ derivative are suppressed by $M_{hi}^{2n}$ after renormalization, i.e.,
\begin{eqnarray}
C^{(\prime)R}_{2n}\sim \frac{4\pi}{M_N M_{hi}^{2n+1}}, \label{eq4}
\end{eqnarray}
where the superscript $R$ denotes the value after renormalization. 
Note that a non-relativistic propagator $G_0$ scales like $\frac{M_N Q}{4\pi}$, with $Q$ the typical c.m. momentum of the nucleon \cite{bira9808007}.
Thus, $C^{(\prime)R}_0 G_0\sim \frac{Q}{M_{hi}}$ and $C^{(\prime)R}_0 G_0 C^{(\prime)R}_0$ are suppressed by $\frac{Q}{M_{hi}}$ with respect to $C^{(\prime)R}_0$.
The resulting scattering amplitude can then be obtained order by order perturbatively. However, in this EFT, no bound state is allowed as bound states require at least part of the 
interaction to be treated non-perturbatively.

It has been shown \cite{pionless} that to describe bound states, rearrangement of the series is necessary. In this second scenario one needs
\begin{eqnarray}
C^{(\prime)R}_{2n}\sim \frac{4\pi}{M_N \mathcal{N}^{n+1}(M_{hi})^{n}}, \label{eq5}
\end{eqnarray}
where $\mathcal{N}<<M_{hi}$ is a low energy scale which can be linked to the existence of bound states. Now, 
$C^{(\prime)R}_{2n}G_0 \sim \frac{Q}{\mathcal{N}}\sim 1$, which means any further iteractions of $C^{(\prime)R}_{2n}$ is of the same order as $C^{(\prime)R}_{2n}$.
This results in an EFT where the LO amplitudes are obtained non-perturbatively from the LO potential $C^{(\prime)R}_0$. 
Starting from NLO, $C^{(\prime)R}_{2n}$ are included perturbatively in the so-called distorted-wave-Born-approximation (DWBA).
This scenario is the standard power counting of pionless EFT.

Note that one can either use dimensional regularization or regulators in the loop calculations. 
Due to the simplicity of the interaction, analytical solution of the loops exists and the final on-shell LO T-matrix can be expressed by a resummation of a geometric series.
Higher-order corrections enter through perturbation theory and all the renormalized LECs can be related to parameters in the effective range expansion. 
After renormalization, RG has been checked explicitly and is found to be well-behaved \cite{pionless}.  

\subsection{Few-body level}
A surprising result is found when one adopts the two-body interaction from pionless EFT---which is well-organized in the NN sector---to calculate the three-particle system. 
One observes the so-called Thomas-collapsing effect \cite{thomas} in numerical calculations. By analyzing the interaction between the
dimer (formed by the first two particles) and the third particle, it can be shown analytically that
the resulting amplitude does not meet the RG requirement \cite{bira99,pionless3,bira_rev}.
A simple intuitive argument is that, the number of pairs ($P_2$) of two-body interactions that appear in an A-body system is:
\begin{eqnarray}
P_2=\frac{A(A-1)}{2},
\end{eqnarray}
while the corresponding appearance of the kinetic term is $A-1$ (one of the kinetic terms goes into the total c.m. of the system). 
Thus, for $A\geq 3$, the NNN system will collapse when $\Lambda\rightarrow \infty$ if the interaction---which is renormalized to produce the NN bound state---is purely attractive\footnote{The interaction 
pairs consist of $v_{12}$, $v_{23}$ and $v_{13}$ in a three-particle system but are only accompanied by
two kinetic terms. The extra pair of the purely attractive interaction causes the system to collapse.}.
The only solution without destroying the two-body power counting is to adopt a repulsive three-body force at LO in the many-body calculations.
Once this is done, the number of three-particle-subsets in the $A>4$ systems---$A(A-1)(A-2)/6$---is always larger than the number of two-body pairs ($A(A-1)/2$).  
It is then quite likely that no higher-body force will be needed at LO in order to have stable results, as repulsive interactions do not require extra boundary condition 
in order to reach RG-invariance. Indeed, calculations suggest that a RG-invariant description for systems up to $A=16$ is achieved \cite{pionless16} at LO and NLO, though the $^{16}$O is found to be not stable
against breakup into four $^4$He.

Once the LO amplitude is calculated, subleading corrections enter perturbatively. 
Power counting of three-body forces at higher order and partial-waves are well-studied within few-body systems in Refs. \cite{pionless3a}. It is shown that, in the strong force sector,
the next three-body force (the one with the lowest momentum-dependence) enters at NNLO\footnote{When Coulomb is included non-perturbatively, there are evidences that a three-body force with new isospin structure 
is needed at NLO \cite{pionless3cou}. The issue regarding how Coulomb can be included is further studied in Ref. \cite{pionless3cou1}.}.    
Moreover, Ref. \cite{pionless4} shows that 
a four-body force is needed already at NLO in order to have RG-invariant systems for more than four particles.
This surprising feature stems from the absence of long-range interactions, which affects the A-body wavefunctions near the origin.
For bosonic system this results in a conjecture \cite{pionless4} that an A-body force will be needed at N$^{A-3}$LO. 

It was demonstrated in Ref. \cite{unitarity} that the nuclear system might be approached from the unitarity limit.
The NN system has scattering lengths much larger than the range of the interaction. For example, the neutron-proton scattering length $a_{np}=-23.7 (5.4)$ fm in the $^1S_0(^3S_1-^3D_1)$ channel,
is much larger than the range $\sim 1/m_{\pi}=1-2$ fm. At the unitarity limit, the scattering length $a\rightarrow \infty$ and therefore the two-body system is scale invariant. 
The remaining parameter that enters at LO is the three-body force. 
This scheme is very attractive as it suggests that, within its range of validity, only one parameter is enough to describe basic properties of many-body systems. 
Carrying out this idea to an extreme, it is shown \cite{denis,denisb,denisc,denisd,bira_eos} that the equation of state of pure neutron matter can be approached from the unitarity limit, where the 
LO is governed by one single parameter---the Bertsch parameter \cite{bertch}.

\subsection{Do we understand power counting in pionless EFT?}
Up to this point, it is clear that the power counting in pionless EFT has been understood fully at least up to the few-body level. 
I list two open-questions as follows:
\begin{itemize}
\item Whether the conjecture that A-body forces will be needed at N$^{A-3}$LO for bosonic system can be proven, and 
what is its impact in fermionic systems?
\item What is the applicability of pionless EFT? At which nuclei it stops to work.
\end{itemize}
Answering the above question would rely mainly on numerical calculations.

\section{Power counting in pionful EFT}

Pions have played the central role and is regarded as the most important building block of the NN interaction since 1930's \cite{Yukawa}. 
The Lagrangian including nucleons and pions as degrees of
freedom was constructed
in the 70's and the resulting chiral perturbation theory \cite{chpt,chptb,chptc,chptd} has been utilized to describe the $\pi \pi$ and $\pi N$ processes quite successfully. 
The first attempt to construct a pionful EFT in the NN sector as advocated by Weinberg \cite{We90} was carried out by van Kolck, \textit{et al.} almost three decades ago \cite{bira,birab}. 
Since then, numerous works have been accomplished to extend the so-called Weinberg counting (WPC) up to very high order \cite{kaiser,n4lo,n4lob,n5lo,Epeln5lo}.
Indeed, when utilized correctly, EFT with explicit pions would be much more powerful than the pionless EFT due to the increase of $M_{hi}$. Unfortunately, a direct implementation of chiral
potential based on WPC generates severe RG problems. The power counting of pionful EFT becomes a topic which is still debated intensively as will be discussed in the next section.

\subsection{NN level}
The chiral Lagrangian is consists of $L_{\pi \pi }$, $L_{\pi N}$ and $L_{NN}$ part. 
$L_{NN}$ has exactly the same form as Eq. (\ref{la}), and 
\begin{eqnarray}
L_{\pi \pi } &=&\frac{1}{2}(\partial _{\mu }\mathbf{\pi }\partial ^{\mu }%
\mathbf{\pi }-m_{\pi }^{2}\mathbf{\pi }^{2})+..., \\ \label{lapi0}
L_{\pi N} &=&\frac{g_{A}}{2f_{\pi }}N^{\dag }(\overrightarrow{S}\cdot \tau
\cdot \overrightarrow{\nabla }\mathbf{\pi })+...,
\label{lapi}
\end{eqnarray}
where $\mathbf{\pi }$ is the pion field, $\overrightarrow{S}$ and $\tau$ are the spin and isospin operators, $g_A\sim 1.27$ is the axial coupling constant and $f_{\pi}\sim 93$ MeV is the pion decay constant. 
Note that $L_{\pi N}$ in Eq. (\ref{lapi}) is obtained by the heavy baryon non-relativistic reduction (HBChPT) \cite{hb,hbb,hbc}. 
A non-relativistic reduction is necessary since the mass of nucleon $M_N\gtrapprox M_{hi}$ and therefore needed to be separated from the 4-momentum to allow the power counting.
Various methods exist to perform the non-relativistic reduction \cite{non-rel,non-relb,non-relc,non-reld}, and already at this level controversies appear as to be discussed later.

Let us continue with the conventional approach (WPC) first, which is to follow the heavy-baryon formalism and calculate the irreducible pion-exchange diagrams order by order.
At the end, one obtains the so-called chiral potential up to a certain order. Then, due to the existence of bound states, WPC prescribes
a full non-perturbative treatment. That is, one iterates the chiral potential to all orders under an ultraviolet cutoff $\Lambda$ in the Lippmann-Schwinger or Schr\"{o}dinger equation to obtain the NN amplitude. 
Note that the potential contains irreducible long-range (pion-exchange) diagrams
truncated at a certain order with contact terms corresponding to the divergence of those diagrams.

Within a certain range of $\Lambda$ (typically $\sim 400-800$ MeV), it is possible to adjust the LECs in WPC up to NNLO (N$^3$LO or higher)\footnote{Here the order is labeled according to the potential under WPC.} to 
obtain reasonable (excellent) fit to the NN scattering data \cite{n4lo,n5lo,Epeln5lo,Epel,Epelb,idaho,idahob,nnloopt,Epelmore}.

Note that the mass difference between $\Delta(1232)$ excitation and nucleon is smaller than $300$ MeV. If $\Delta(1232)$ is not included explicitly in the loop calculations (e.g., 
in two-pion-exchange diagrams), one is in the risk of having a radius of convergence smaller than
the estimated $M_{hi}$ ($\sim600$ MeV). Therefore, in principle one should include the $\Delta$ in the EFT to recover
its full power. Since its first evaluation in Ref. \cite{bira}, Delta-full potential has been refined up to N$^3$LO  \cite{delta,deltab} and applied within WPC to several calculations \cite{delta_ab,delta_ab2,delta_ab3}. Ref. \cite{nnlodelta} indicates that a better description
of nuclear data, in particular, the saturation point can be achieved with the Delta-ful potential.

Despite the phenomenological success in terms of describing data, two problems appear in WPC already at LO. First, as pointed out in Refs. \cite{ksw,ksw2,ksw3}, once the one-pion-exchange potential (OPEP) is iterated non-perturbatively there is no way to properly renormalize
the divergence caused by varying the pion mass. This issue motivates the so-call KSW counting, which treats the pion-exchange perturbatively \cite{ksw,ksw2,ksw3}. However, it was shown later that 
this counting suffers from convergence problem, in particular at the spin-triplet channels \cite{fleming}\footnote{Note that a recent work \cite{k} shows that the problem of KSW only persists in the $^3$S$_1$-$^3$D$_1$ and
$^3$P$_0$ channels by performing higher-order calculations.}.  
The second problem, as pointed out in Ref. \cite{nogga}, states that even without considering the chiral extrapolation problem\footnote{I.e., problems involving varying the pion mass. }, WPC 
lacks of necessary contact terms (in singular and attractive higher partial-waves ($l\ge 1$)) to achieve RG-requirement.
Later it was pointed out that the same problem (the lacks of RG) exists at every order of WPC \cite{Ya09A,Ya09B,ZE12}, due to a Wigner-bound like effect \cite{wigner,wigner2}.

As a matter of fact, WPC fails to satisfy RG-requirement and is therefore subjected to the danger of becoming a model.
However, this conclusion is in conflict with the traditional approach of nuclear physics, which absorbs physics into parameters in an effective potential and then solves the amplitudes non-perturbatively.
In this spirit, ``no" EFT potential with more than one LEC (per partial-wave)\footnote{Under 
the condition that LECs are expressed in power of momentum as listed in Eq. (\ref{eq3}).} can satisfy the RG-requirement, due to the occurrence of Wigner-bound like effect. 
Therefore, a general attitude is to limit $\Lambda<M_{hi}$ and assume that the power counting organized at potential level will survive
through a ``moderate" iterations to the final observables. Ref.  \cite{ge} further argued that adopting a cutoff $\Lambda>M_{hi}$ will cause the ``peratization"
of the resulting amplitudes.
This idea is illustrated in Ref.  \cite{ge1} via a pionless example as follows.
Consider the scenario that the first two terms in Eq. (\ref{eq3}) needed to be iterated to all order.
To label the number of iterations of V, one can insert a parameter $\hbar$ in the Lippmann-Schwinger equation, i.e.,
\begin{equation}
T=V+\hbar \,V GT,
\label{one}
\end{equation}
so that $VGV\sim \hbar$, $VGVGV\sim \hbar^2$, ..., etc. 
Then the $^1$S$_0$
amplitude can be expressed as \cite{pionless1}
\begin{align}
&T_{\rm NLO}(q) = \notag \\
&\frac{c_2 \left[\hbar\, c_2 \left(I_3
   q^2-I_5\right)-2
   q^2\right]-c}{\hbar \, I\left(q^2\right)
   \left[c_2 \left( \hbar\,c_2
   \left(I_5-I_3 q^2\right)+2
   q^2\right)+c\right]-\left(\hbar I_3
   c_2-1\right){}^2},\label{1s0ampl1}
\end{align}
where, under cutoff $\Lambda$,
\begin{align}
I_n&= - m\int \frac{d^3k}{(2\,\pi)^3}\ k^{n-3}\,\theta(\Lambda-k)
=-\frac{m\,\Lambda^n}{2\,n\,\pi^2}\,,\nonumber\\
I(p^2)&= m\int \frac{d^3
k}{(2\,\pi)^3}\,\frac{1}{p^2-k^2+i\,0^+}\,\theta(\Lambda-k) \notag \\
&= -\frac{i\,p\ m_N}{4 \pi}-\frac{m}{2\,\pi^2}\left[ \Lambda
+\frac{p}{2}
\,\ln\frac{1-\frac{p}{\Lambda}}{1+\frac{p}{\Lambda}}\right]\nonumber\\
& =-\frac{i\,p\,m}{4 \pi}-\frac{m\,\Lambda}{2\,\pi^2} +\frac{m
\,p^2}{2 \,\pi^2 \Lambda}+O\left(\frac{1}{\Lambda^2}\right) \label{loopintegrals}. 
\end{align}
Ref.  \cite{ge1} then claims that renormalization is only achieved when the divergence of each diagram in the infinite series are removed by
their corresponding counter terms. In order to achieve that, one needs to promote infinitely many counter terms with higher derivatives to renormalize $T_{NLO}(q)$.
Ref.  \cite{ge1} then shows that, by doing this, the resulting amplitude satisfies ``perturbatively renormalizable" condition\footnote{I.e., after renormalization, there is an one-to-one correspondence 
between the expanded (in $\hbar$) resummed series (Eq. (\ref{1s0ampl1})) and the perturbative diagrams.} and is free of the Wigner-bound problem, 
while the renormalization performed in Ref.  \cite{pionless1} does not. 
Thus, one should either performs the renormalization as done in Refs.  \cite{ge1,ge_nonrel,ge_nonrel2,ge_nonrel3} or keep the cutoff low
to avoid ``peratization".

However, in an EFT one should not take any a priori assumption to assume that a particular treatment of an interaction (in this case, a non-perturbative treatment of $c+c_2$)
will result in an amplitude satisfying the EFT power counting. It could be that an incompleted higher-order effect is generated due to the iteration, and one should either expand and truncate the result properly or change the 
interaction itself and then perform the renormalization.
In the above example, it is clearly shown \cite{bira99} that the resulting amplitude (Eq. (\ref{1s0ampl1})) should be expanded up to $q^2$, and with $c_2$ enters perturbatively through the distorted-wave-Born-approximation (DWBA)\footnote{See, e.g., Ref. \cite{ge_nonrel} for alternative opinion.}. Then, one
performs renormalization of the two lowest order terms ($q^0$ and $q^2$ terms) to the effective range expansion. All other effects are of higher-order. If a proposed power counting is wrong,
forcing a removal of divergences in each individual diagram (by introducing additional contact terms not prescribed before) will do nothing good but just hide the problem---which originally might be easily detected
by a simple RG-check.  
The problem will still be revealed finally by a Lepage-plot-like analysis. 
In this case, Eq. (\ref{eq5}) shows that the renormalized $c^R_{2n}$ is of the same order
of $(c^R_2)^{n}$, but is not included in Eq. (\ref{1s0ampl1}). Therefore, before expansion, Eq. (\ref{1s0ampl1}) contains incompleted higher-order effects and the resulting amplitude is
wrong anyway regardless how the renormalization is done. 

Note that when pions are presented, forcing the same removal of divergences in
all diagrams will introduce incompleted higher-order effects, as 
a function formed by all higher-order contact terms is introduced to absorb the divergences. This
can destroy or improve the agreement of the resulting amplitude with respect to the prescribed power counting, but in an uncontrollable way.
If, at a certain order, the short range physics supposed to enter cannot be represented sufficiently by a simple combination of contact terms,
one could enrich the structure of contact terms by introducing a field re-definition to incorporate auxiliary fields such as the dibaryon or others, so that the power counting of both the long- and short-range physics remains in a tractable manner.

On top of ``perturbatively renormalizable",
Refs. \cite{ge_nonrel,ge_nonrel1} advocate an alternative 
procedure \cite{add} to introduce additional terms\footnote{Symmetry-preserving higher-derivative terms are introduced in the effective Lagrangian of baryon chiral perturbation
theory.} into the relativistic Lagrangian and obtain the propagator and pion-exchange potential in an alternative form. 
As a result, the divergences of the iterated diagrams are greatly reduced.
Then, the interaction is treated non-perturbatively with the ``perturbatively renormalizable" scheme applied to the resulting T-matrix.

However, as mentioned before, due to the large nucleon mass in the four-momentum one cannot discriminate the importance between propagators or vertices generated by the
higher- and lower-derivative terms in the Lagrangian.
For individual diagrams, it was demonstrated \cite{add} that one can perform calculations directly in the relativistic form and then apply appropriate 
expansion later to obtain results which match HBChPT up to the relevant order\footnote{See also Refs. \cite{Lisheng,Lisheng2,Lisheng3,Lisheng4,Lisheng5,Lisheng6,Lisheng7} for further studies following 
this direction and the comparison of results between HBChPT and the covariant framework.}. 
Meanwhile, before a proper expansion, each diagram contains incompleted higher-order contributions due to the relativistic treatment. 
Thus, on top of the potential problem of forcing the ``perturbatively renormalizable" condition on the non-perturbative treatment, additional error could be generated.
In particular, the pion-exchange part of OPEP obtained in this way behaves as
\begin{eqnarray}
V_{\pi}(p\rightarrow \infty,p^{\prime}\rightarrow \infty)\sim \frac{1}{pp^{\prime}} \label{rel3p0}
\end{eqnarray}
in the $^3$P$_0$ channel. Thus, in contrast to the usual non-relativistic OPEP, the above potential is non-singular and therefore does not require a contact term to achieve RG-invariance.  
Promoting a contact term to LO in the non-perturbative treatment will destroy the RG \cite{Ya09A}, unless extra care is taken to further subtract the divergences.
This is demonstrated in the ``non-perturbatively renormalized" versus ``subtractively renormalized" phase shifts in Fig. 4 of Ref.  \cite{ge_nonrel1}.

Power counting with resulting amplitudes converge with respect to cutoff (but not necessary satisfy ``perturbatively renormalizable" condition) exists in three versions \cite{Birse,Valdper,BY}.
Though differ in some aspects, all of them treat subleading interaction perturbatively. The LO interaction consists of OPEP with appropriate contact terms (to ensure RG-invariance)
and is treated non-perturbatively. At this order, a promotion of contact terms (with respect to the LO WPC) is required for those singular and attractive channels
to ensure that the boundary condition is fixed \cite{nogga}.
Recent studies also suggested that all $l\geq 1$ partial-waves except $^3$P$_0$ might enter perturbatively due to the effect of the centrifugal barrier \cite{vald17,bingwei18}.
Once a promotion at the LO is required, all higher-order contact terms which enter perturbatively are promoted at the same time, due to a peculiar 
structure in the distorted wave near the origin.
Note that the entrance of additional scale $M_{lo}$ analog to $\mathcal{N}<<M_{hi}$ in Eq. (\ref{eq5}) is presented in the new power counting.
For example, for singular and attractive P-waves where contact terms need to be promoted, the amplitudes scale as $\frac{(p_{cm},p_{typ})^3}{M_{lo}^3}$,
$\frac{(p_{cm},p_{typ})^{5}}{M_{lo}^3 M_{hi}^2}$ and $\frac{(p_{cm},p_{typ})^{6}}{M_{lo}^3 M_{hi}^3}$ at LO, NNLO and N$^3$LO according to the power counting proposed in Ref.  \cite{BY}. 

Finally, special treatments might be necessary for the $^1$S$_0$ channel as there is a large discrepancy between the LO phase shifts obtained through WPC, KSW, or modified power counting \cite{nogga,Birse,Valdper,BY}
and the Nijmegen analysis \cite{nij}. Studies \cite{DB,Bs,1s0d} suggested that adopting the auxiliary dibaryon field together with OPEP could provide 
significant improvement to the LO amplitude. This improvement could become crucial in many-body calculations as will be addressed in the next section.

In summary, despite the phenomenological success of WPC and many studies toward the improvement, the power counting in pionful sector is still much 
less understood with respect to the pionless case. Nevertheless, a general framework toward a RG-invariant power counting has been laid out. 
Since RG-invariance (as defined previously in footnote 3) is just the minimum requirement of an EFT \cite{nogga}, a detailed analysis of power counting utilizing Eq. (\ref{eq2}) is required and is on-going \cite{jerry_lepage}.

\subsection{Beyond NN-level}

WPC has been applied widely to nuclear structure calculations in the few- and many-body sector. It has been shown that together with three-body forces
and a more restricted $\Lambda$ ($\sim 400-500$ MeV), binding energies and radii of nuclei can be reasonably reproduced \cite{abn3lo,abn3loa,abn3lob,abn3loc,abn3lod,abn3loe,abn3lof,abmore,abmoreb,abmorec,abmored,abmoree,abmoref}.
To get the most out of WPC, it is shown that one could perform a general fit of LECs to a wider range of nuclear properties to achieve a better description of many-body systems \cite{nnloopt,nnlosat}.
It is shown that an even better description
of nuclear data can be achieved with the Delta-full potential \cite{nnlodelta}.

However, not all observables can be well-described by WPC. In the few-body level, there exists the so-called $A_y$ puzzle, i.e., the
nucleon vector analyzing power in elastic deuteron-nucleon scattering below 30-MeV laboratory energy is not reproduced
by WPC and other phenomenological potentials \cite{Ay,Ay2,Ay3,Ay4,Ay5,Ay6,Ay7}. In the intermediate mass nuclei, there exists systematic overbinding \cite{overbind}
and a ``radius problem" \cite{radius}. Note that the sources of the above problems are still unclear at current stage, and might not be directly related to the problem of power counting. 
Moreover, several interactions inspired by Chiral EFT are able to describe radii and binding energies from light to heavy nuclei better, see Ref. \cite{radius1} for a recent review of this issue. 

\begin{figure}[h]
\includegraphics[width=8cm]{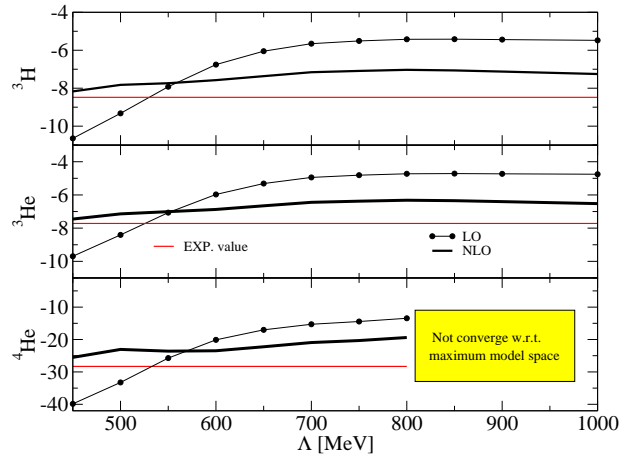}
\caption{Binding energy of $^3$H, $^3$He and $^4$He at LO (black-circle-line) and up to NLO (black solid-line) versus cutoff. For $^4$He, trustworthy results can only be
obtained up to $800$ MeV due to computational limit.}
\label{fig1}
\end{figure}

On the other hand, there are only a handful calculations based on power counting other than WPC. 
RG-invariant results for A=3 systems are obtained at LO \cite{nogga} and up to NLO \cite{song}
based on one version of the RG-invariant power counting \cite{BY}. 
In a recent work \cite{ours}, the binding energies of $^3$H, $^3$He and $^4$He are calculated according to 
the power counting proposed in Refs.  \cite{BY}. The results as a function of $\Lambda$ are presented in Fig. \ref{fig1}.
As one can see, reasonable and RG-invariant
binding energies can be obtained by including just up to NLO contributions in the new power counting scheme. 
However, the same interaction fails to produce an $^{16}$O more bound than four $\alpha$ particles. Thus, although the power counting seems to work fine for A$\leq 4$ systems, the A=16 pole structure is not correctly reproduced.
Since subleading interactions enter perturbatively in the new power counting, it is not clear whether the wrong pole structure can be corrected in a perturbative way.
A promotion of three-body force to LO for heavier nuclei is likely to be the solution. 

\subsection{An $A$-dependent scenario for the higher-body forces}

Now I discuss the possible scenario regarding the promotion of higher-body forces. So far, the power counting of a higher-body force 
is decided mainly based on either the RG-analysis or NDA. After the power counting is verified up to a certain order based on few-body calculations (say, up to A-particle systems),
a common expectation is that higher-body forces which are not required in smaller systems will
not appear in the calculations of heavier systems. This expectation can, however, be wrong. One naive estimation of the importance of higher-body force can be done by simply
counting the number of pairs for an N-body interaction. Table \ref{t1} lists the occurrence of the two- and three-body force in an A-body system. As one can
see, the number of triplets grow as $\sim A^3/6$ for larger A and exceeds the number of doublets $\sim A^2/2$ after $A>5$. Thus, if the relative strength of the triplet versus
doublet is not suppressed by more than $3/(A-2)$, both of them would need to be included in the A-particle system calculation.   
This means, as the increase of A, many-body forces will eventually needed to be promoted\footnote{Pauli principle could weaken (kill) the above effect for some of the long- (short-) range 
higher-body forces, but there are strong evidence from studies of nuclear matter
equation of state that it is necessary to adopt either a three-body or a density-dependent two-body term already at LO in order to describe the empirical data \cite{skyrme,skyrme2,skyrme3,skyrme4}.}. 
\begin{table}[bt]
\begin{tabular}{|c|c|c|}
\hline
  & number of doublet & number of triplet \\
A & $\frac{A(A-1)}{2}$ & $\frac{A(A-1)(A-2)}{6}$ \\
\hline
3 & 3 &1 \\
\hline
4 & 6 &4 \\
\hline
5 & 10 &10 \\
\hline
6 & 15 & 20 \\
\hline
\end{tabular}%
\caption{Number of double and triplet in an A-particle system.}
\label{t1}
\end{table}

\section{Summary}

The EFT approach to low energy nuclear
physics allows
one to build inter-nucleon interactions based on the symmetries of QCD at
low energy. When the power counting is fully understood, the interaction can be
considered as the low energy expansion of QCD. 
In this regard, at least up to few-body level, the power counting is well understood in the pionless sector.
The remaining open questions concern mainly the power counting of many-body forces and the range of applicability in the nuclear structure aspect are to be studied numerically. 
On the other hand, despite extensively studies, the power counting in pionful EFT remains less understood.


\section{acknowledgments}
I thank A. Ekstr\"om and U. van Kolck for
useful discussions and suggestions. Computations were performed on resources provided by the
Swedish National Infrastructure for Computing (SNIC) at NSC, Linkoping.

%

\end{document}